\documentclass[aps,prb,groupedaddress,amsmath,amssymb,twocolumn]{revtex4}

\usepackage{amssymb}
\usepackage{amsmath}
\usepackage{graphicx}
\usepackage{subfigure}
\usepackage{textcomp}
\usepackage{color}
\usepackage{amsfonts}
\usepackage{bbold}
\usepackage{dsfont}
\usepackage{epsfig}
\usepackage{hyperref}

\begin{document}
\title{Class of Rashba ferroelectrics in hexagonal semiconductors}

\author{Awadhesh Narayan}
\email{awadhesh@illinois.edu}
\affiliation{Department of Physics, University of Illinois at Urbana-Champaign, Urbana, Illinois, USA.}

\date{\today}

\begin{abstract}
We present a class of Rashba systems in hexagonal semiconducting compounds, where an electrical control over spin-orbital texture is provided by their bulk ferroelectricity. Our first-principles calculations reveal a number of such materials with large Rashba coefficients. We, furthermore, show that strain can drive a topological phase transition in such materials, resulting in a ferroelectric topological insulating state. Our findings can open avenues for interplay between Rashba effect, ferroelectricity and topological phenomena.
\end{abstract}

\maketitle

\textit{Introduction--} In inversion-asymmetric systems with spin-orbit coupling, the degeneracy of energy bands is lifted and they become spin split. This is the celebrated Rashba effect~\cite{rashba1960properties}, which has traditionally been intensively investigated in two-dimensional surfaces and interfaces~\cite{lashell1996spin,nitta1997gate}. More recently a bulk Rashba effect has been found in the layered material BiTeI~\cite{ishizaka2011giant}.

\begin{figure}[b]
\begin{center}
  \includegraphics[scale=0.55]{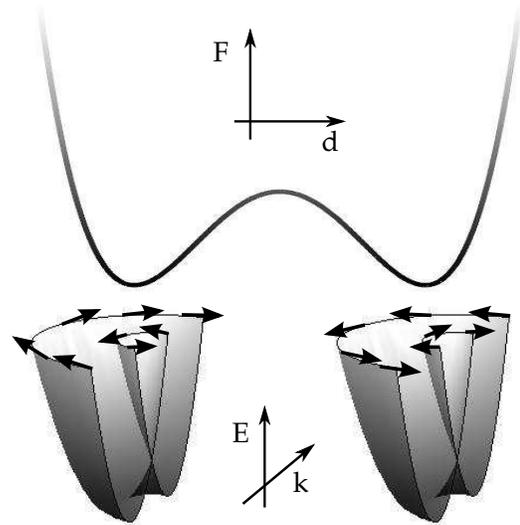}
  \caption{Schematic illustrating the principle of ferroelectric Rashba effect. Free energy of the ferroelectric is shown as a function of the distortion from the centrosymmetric structure. When polarization of the material is switched, the underlying Rashba-like spin texture (here shown superimposed with the energy dispersion) can also be reversed between the two minimum energy configurations. This gives an electrical handle over spin degrees of freedom.}  \label{rashba_schematic}
\end{center}
\end{figure}

Among the materials forming in inversion-symmetry-breaking structures, ferroelectrics are an interesting subclass. They exhibit a spontaneous electric polarization, which can be switched by an external electric field. One would expect the lack of inversion symmetry, in this class of materials, when combined with large-enough spin-orbit coupling, to lead to interesting Rashba ferroelectric behavior. Indeed, recently this combination has been shown in GeTe to lead to a giant Rashba effect with full control over spin texture via electric field control of the polarization~\cite{di2013electric}. We schematically show the underlying principle of the ferroelectric Rashba effect in Fig.~\ref{rashba_schematic}. As one applies an electric field to switch between the two ferroelectric states (two minima on the free energy surface), the spin texture of the Rashba bands is also reversed at the same time. This remarkable coupling between spin and electric dipoles can provide an arena for as-yet-unexplored fundamental physical effects. From an applications perspective, these materials could be utilized in spintronic devices. This electrical control knob over the spin degrees of freedom, can be effectively exploited, for instance, in a modified Datta-Das spin transistor~\cite{picozzi2014ferroelectric}. Despite such promise of novel physical effects and applications, only a handful of Rashba ferroelectric materials are known, aforementioned GeTe and three halide perovskites~\cite{kim2014switchable}.

Topological insulators are another class of materials that have caught wide attention in recent years~\cite{hasan2010colloquium,qi2011topological}. These large spin-orbit materials have an insulating bulk, while at the same time having topologically protected metallic surface states. In these surface states the electron spin is coupled to its momentum, and their existence is guaranteed by underlying symmetries of the bulk. Since, both ferroelectric Rashba materials and topological insulators have spin-orbit interaction as a common ingredient, it would be exciting to find materials which can exhibit physics of both. A ferroelectric topological insulator could also lead to intriguing electrical control over the topological surface states.

\begin{figure*}[t]
\begin{center}
  \includegraphics[scale=0.6]{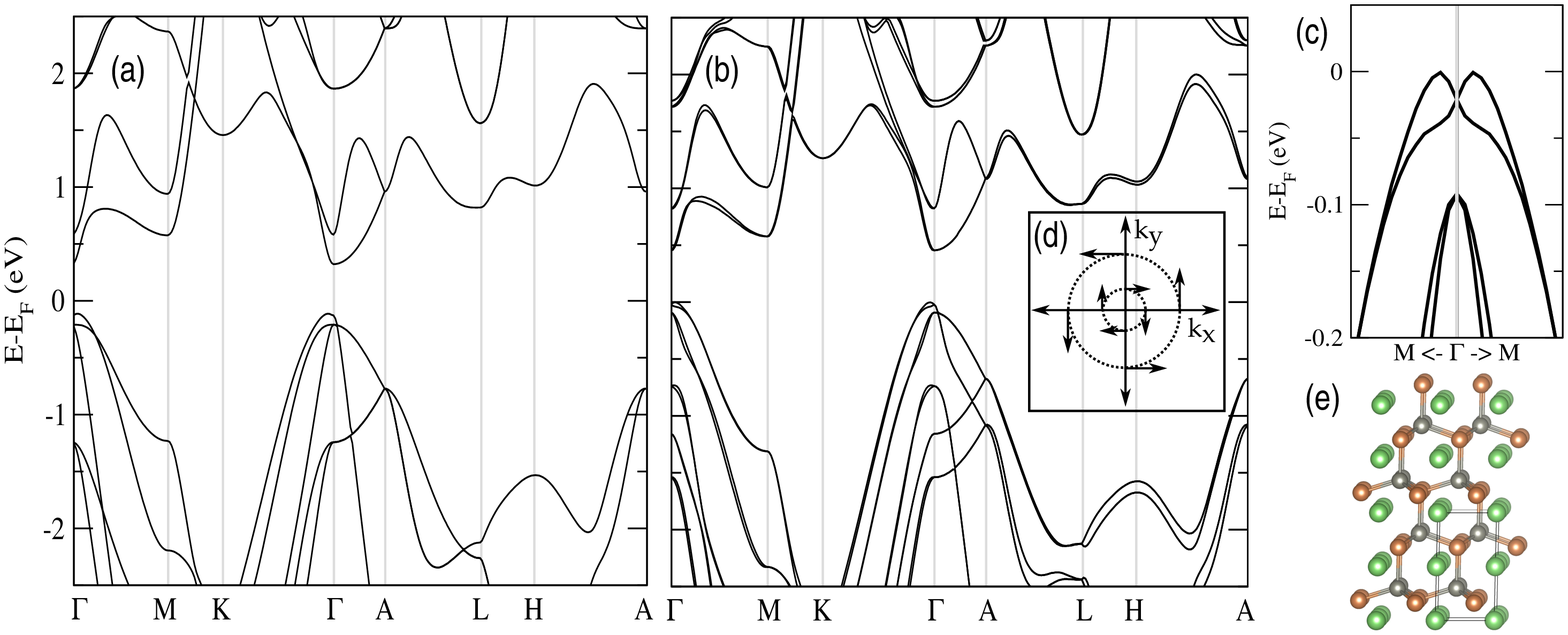}
  \caption{(Color online) Band structure of LiZnSb along high-symmetry directions in the Brillouin zone (a) without spin-orbit and (b) including spin-orbit interaction. Note the lifting of degeneracy of the bands upon inclusion of spin-orbit coupling, in particular close to $\Gamma$. (c) A zoom close to the Fermi energy around $\Gamma$ reveals characteristic Rashba-split bands. (d) Inset shows the in-plane spin texture around the center of the Brillouin zone plotted at an energy of 0.035 eV below the Fermi level. The out-of-plane spin component is negligible. (e) LiGaGe parent structure type (space group P6$_{3}mc$) of the compounds studied in this work. Black lines indicate the unit cell. }  \label{bands_pristine}
\end{center}
\end{figure*}

A close examination of the known Rashba ferroelectrics reveals that they share some common features, they are relatively small gap materials and they are comprised of heavy elements. The heavy atoms provide the necessary spin-orbit coupling, while a small gap affords efficient coupling through the spin-orbit interaction. Guided by these considerations, we have carried out a search for ferroelectrics with large Rashba coefficients. While heavy elements are present in a number of known ferroelectrics, these materials inevitably have a large band gap, which hinders the Rashba effect. A notable exception is a class of ferroelectrics in LiGaGe structure type proposed by Bennett and coauthors~\cite{bennett2012hexagonal}. Not only do many of them have heavy atoms like Bi or Sb, a number of them have moderate band gaps of less than an electron-volt. In this contribution we show that hexagonal semiconductors in LiGaGe structure type are a promising class of Rashba ferroelectrics. Our first-principles calculations reveal that these have large Rashba coefficients, which are an order of magnitude greater than that obtained in traditional Rashba surfaces and interfaces. We add several promising candidates to the family of ferroelectric Rashba materials. Furthermore, we show that a strain can drive a topological phase transition in this class of materials. This results in a ferroelectric topological insulating state. Such a confluence of ferroelectricity and topological states can potentially open avenues for interplay between seemingly quite different physical phenomena.

\textit{Computational Methods--} Our first-principles density functional theory calculations were performed using the Quantum-Espresso code~\cite{giannozzi2009quantum}. We used the Perdew-Burke-Ernzerhof (PBE) parametrization to the exchange-correlation functional~\cite{perdew1996generalized}. An energy cutoff of 40 Ry was chosen for the plane-wave basis, along with a $6\times 6\times 6$ Monkhorst-Pack $k$-point mesh. We started with the structural parameters reported in Ref.~\onlinecite{bennett2012hexagonal}, and optimized the geometries. The compounds that we have identified crystallize in LiGaGe structure type in the $P6_{3}mc$ polar space group. This structure, shown in Fig.~\ref{bands_pristine}(e), can be considered as the non-centrosymmetric variant of the inversion-symmetric $P6_{3}/mmc$ structure, which does not have a buckling of the atomic planes.

\textit{Results and discussion--} We first present a detailed analysis of LiZnSb, a representative of this class of Rashba ferroelectrics. Figure~\ref{bands_pristine}(a) shows the band structure of LiZnSb without spin-orbit coupling. It is a narrow gap semiconductor, with the valence band maximum and conduction band minimum lying at the $\Gamma$ point. On including spin-orbit coupling [Fig.~\ref{bands_pristine}(b)], a number of band degeneracies are lifted, except along the $\Gamma-A$ line where it is suppressed. This lack of degeneracy is specially pronounced in the valence band maximum, both along $\Gamma-M$ as well as $\Gamma-K$ directions, where the two-bands are nearly isotropically spin split. The valence bands are primarily comprised of Sb $p$ states, while the conduction band derives from a combination of Zn and Sb $p$ orbitals. A zoom of the energy bands shown in Fig.~\ref{bands_pristine}(c) reveals a characteristic Rashba-like band structure, where the splitting is much larger in the valence bands than the conduction band.

\begin{table}[b]
\caption{A summary of the Rashba energy splitting ($E_{R}$) in the valence band, the momentum offset ($k_{R}$) and the Rashba coefficient ($\alpha$) along $\Gamma-M$ direction for the hexagonal semiconductors.}
\centering
\begin{tabular}{c c c c}
\hline \hline
Compound & $E_{R}$ (meV) & $k_{R}$ (1/\AA{}) & $\alpha$ (eV\AA{}) \\[0.5ex]
\hline \hline
KMgSb & 10 & 0.024 & 0.83 \\
LiZnSb & 21 & 0.023 & 1.82 \\
LiBeBi & 24 & 0.026 & 1.84 \\
NaZnSb & 31 & 0.024 & 2.58 \\
LiCaBi & 32 & 0.035 & 1.82 \\
\hline \hline
\end{tabular}\label{table1}
\end{table}

We compute the expectation value of the spin operators $\langle S_{\alpha}(k)\rangle=\langle \psi(k)|\sigma_{\alpha}|\psi(k)\rangle$, where $\psi(k)$ are the spinor wave functions, $\sigma_{\alpha}$ are the Pauli spin matrices and $\alpha=x,y,z$. A constant energy cut, which is isotropic around $\Gamma$, is shown at 0.035 eV below the Fermi level in Fig.~\ref{bands_pristine}(d). The in-plane spin texture is indicated by arrows, while the out-of-plane component is nearly zero. The bands show the characteristic Rashba-like spin texture with the two bands having opposite orientations of the spin, the inner band has a clockwise spin texture, while the outer band shows a counter-clockwise rotation of the spins as one goes around the Brillouin zone center. This spin texture can be reversed by switching the polarization of LiZnSb, i.e., by switching buckling of the atomic planes, as shown in the schematic in Fig.~\ref{rashba_schematic}. Thus, we have an electrical way to control and manipulate the spin texture in these materials.

\begin{figure}[t]
\begin{center}
  \includegraphics[scale=0.70]{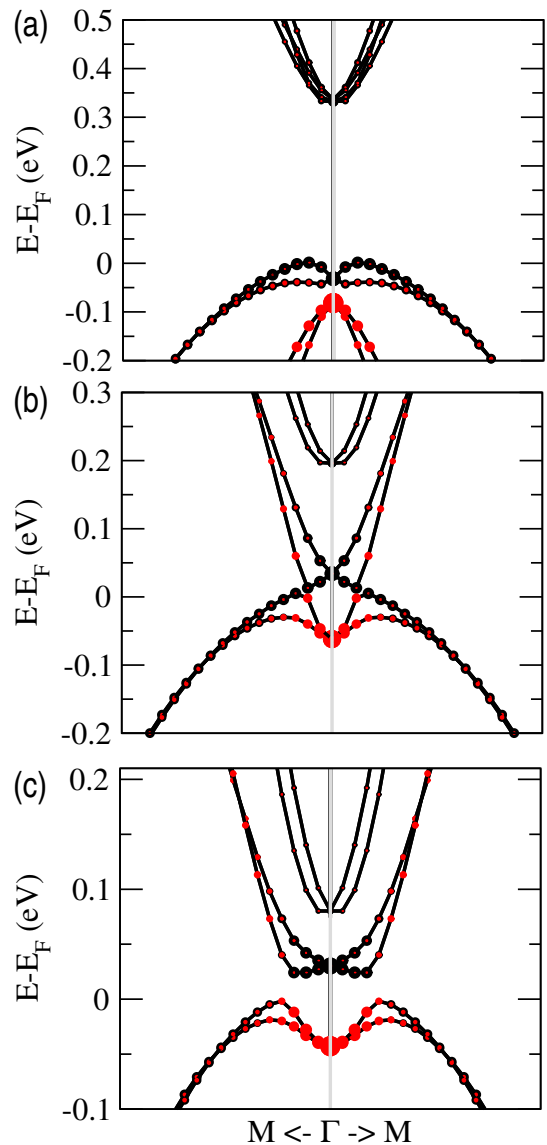}
  \caption{(Color online) LiZnSb under strain. Band structures for (a) $a$=4.6 \AA{}, (b) $a$=4.8 \AA{} and (c) $a$=5.0 \AA{}. The band gap decreases with increasing in-plane lattice constant, with the system transforming into a zero-gap semimetal for $a$=4.8 \AA{}. With further increase in $a$, the gap reopens with an inverted band order around $\Gamma$. Black circles represent the contribution of $j=1/2$ states, while red (grey) circles indicate the contribution coming from $j=3/2$ states.}  \label{bands_pressure}
\end{center}
\end{figure}

In addition to LiZnSb, we have found several other promising hexagonal Rashba ferroelectrics. We summarize their calculated Rashba parameters in Table~\ref{table1}. For all the materials the obtained Rashba coefficients are an order of magnitude larger than those on the surface of a typical large-spin orbit noble-metal like Au(111)~\cite{lashell1996spin}. While the Rashba energy splitting, $E_{R}$, in our proposed set of materials is smaller than that in GeTe and BiTeI by roughly a factor of ten, the momentum off-set between the band degeneracy point ($\Gamma$) and band maximum, $k_{R}$, is also smaller by nearly the same factor. As a consequence, the Rashba coefficient, $\alpha=2E_{R}/k_{R}$, which is the key quantity determining the strength of the Rashba effect, becomes comparable to known bulk Rashba compounds BiTeI~\cite{ishizaka2011giant} and GeTe~\cite{di2013correction}. This compensating feature between energy splitting and momentum offset to yield a large Rashba coefficient is also shared by halide perovskites~\cite{kim2014switchable}. Some of the ferroelectrics in this class of hexagonal semiconductors suffer from the drawback of having a large energy barrier between the polar and reference non-polar state. Of the ones that we have found to show excellent Rashba effect, three compounds have reasonable energy barriers to permit switchability: KMgSb (0.08 eV), NaZnSb (0.16 eV) and LiCaBi (0.01 eV)~\cite{bennett2012hexagonal}. These values of energy barrier are in favorable comparison to that of typical ferroelectrics PbTiO$_{3}$ (0.2 eV) and BaTiO$_{3}$ (0.02 eV). With the largest Rahsba coefficient among this class of compounds, we find that NaZnSb appears particularly promising. It has a large polarization of 0.49 C/m$^{2}$, along with a low barrier for switching (0.16 eV). Out of the listed promising Rashba ferroelectrics, KMgSb, LiZnSb and NaZnSb have a direct band gap. This narrow direct gap feature may also offer opportunities for spin-optical applications.

Given the large spin-orbit strength and small band gaps of our proposed hexagonal Rashba ferroelectrics, a natural question arises: can these compounds be engineered to create topological insulators, which also arise in materials with large spin-orbit interaction? Our calculations reveal the answer to this question in the affirmative. Indeed, in these hexagonal semiconductors a suitable strain can trigger a reduction in band gap, its closure and subsequent re-opening with an inverted band order. In Fig.~\ref{bands_pressure}, we show the evolution of the band structure of LiZnSb with increasing in-plane lattice constant, $a$. At about 8\% change in $a$ ($a$=4.8 \AA{}), the gap closes, and passing through this zero-gap semimetallic state, the gap opens with inverted band character around the $\Gamma$ point. Before closing of the band gap, the main contribution to the top of the valence band comes from $j=1/2$ states (shown in black circles in Fig.~\ref{bands_pressure}). After the gap is re-opened, close to $\Gamma$, the $j=3/2$ states predominantly contribute to the valence band, while $j=1/2$ states now comprise the bottom of the conduction band. This inverted band character signals a topological phase transition from a normal insulator to a topological one, in a manner very similar to canonical topological insulators in the bismuth selenide family~\cite{zhang2009topological,xia2009observation}. The inversion-symmetry-breaking ferroelectric structure continues to be energetically more favorable than the symmetric one throughout this process. This means that in this system we have obtained a topological insulator which is also a ferroelectric at the same time. One of the implications of this finding is that the helicity and spin texture of the topological surface states, forming in a slab geometry, would be controllable by the polarization of the bulk ferroelectric. One could envisage some interesting scenarios which can be created by this coupling. For instance, it may be possible to create a junction between surface states of opposite helicity by having regions with opposite applied electric fields within the same material, which would lead to a gapless interface state~\cite{de2013gapless,habe2013robustness}. By using only electrical control, one would be able to move this junction as well as the interface channel, essentially at will. It also seems to be plausible that such ferroelectric topological insulators could be used to augment traditional ferroelectric devices, like the ferroelectric field effect transistors or tunnel junctions~\cite{bibes2012nanoferronics}, with the added spin degrees of freedom.

Admittedly, our proposal requires a relatively large strain, which may be difficult to achieve using traditional substrate engineering. Nevertheless, we have shown that in principle, one can obtain ferroelectricity and topological phases in the same material. We are hopeful that this would galvanize future search for such multifunctional materials, perhaps even without need for strain. Here we have found a time-reversal invariant topological insulator state coexisting with a ferroelectric. However, more generally one may suspect that there exist materials where topological phases coexist with other kinds of ferroic order, for instance ferroelastic~\cite{salje2012ferroelastic} topological insulators or time-reversal and inversion symmetry broken topological systems with ferrotoroidic order~\cite{spaldin2008toroidal}. One may speculate, that there can also exist topological phases, which harbor multiple ferroic orders. A search for such topological-multiferroic materials will be definitely rewarding, both as playgrounds for fundamental physical phenomena as well as for diverse applications. 

We note that recently a preprint by Liu \textit{et al.} appeared showing that a ferroelectric topological insulator could be obtained in CsPbI$_3$ under a compressive strain~\cite{liu2015strain}.

\textit{Conclusions and summary--} In conclusion, we have found a class of Rashba ferroelectrics in hexagonal semiconductors crystallizing in $P6_{3}mc$ polar space group. Based on the guiding principles of a small energy gap along with presence of heavy atoms, in combination with our first-principles calculations, we identified a promising set of materials with substantial Rashba coefficients, which are an order of magnitude larger than usually studied surfaces and interfaces. The inherent bulk ferroelectricity allows a full electrical control over the spin texture of these Rashba materials. We have also shown that a strain can trigger a topological phase transition in these compounds, resulting in ferroelectricity coexisting with a topological insulating state. Such multifunctionality in our proposed materials can prove to be an interesting arena for interaction between Rashba effect, ferroelectricity and topological physics. 

\textit{Acknowledgments--} I would like to thank Victor Chua for numerous enlightening discussions. I would also like to acknowledge interactions with Domenico Di Sante, Silvia Picozzi and Stefano Sanvito, which partly motivated this work.


%

\end{document}